\documentclass[conference]{IEEEtran}
\IEEEoverridecommandlockouts
\usepackage{cite}
\usepackage{amsmath,amssymb,amsfonts}
\usepackage{algorithm}  
\usepackage{algorithmicx}  
\usepackage{algpseudocode} 
\usepackage{graphicx}
\usepackage{textcomp}
\usepackage{xcolor} 
\usepackage{subfigure}
\usepackage{comment}
\usepackage{multirow}
\usepackage{threeparttable}
\usepackage{booktabs}

\usepackage{tikz}
\usepackage{url}

\def\BibTeX{{\rm B\kern-.05em{\sc i\kern-.025em b}\kern-.08em
    T\kern-.1667em\lower.7ex\hbox{E}\kern-.125emX}}
\begin{document}

\title{ AutoFlow: Hotspot-Aware, Dynamic Load Balancing for
  Distributed Stream Processing}
\author{\IEEEauthorblockN{Pengqi Lu\textsuperscript{$\ast\dagger$}, Liang Yuan\textsuperscript{$\ast$}, Yunquan Zhang\textsuperscript{$\ast$}, Hang Cao\textsuperscript{$\ast\dagger$}, Kun Li\textsuperscript{$\ast\dagger$}}
\IEEEauthorblockA{\textsuperscript{$\ast$}\textit{State Key Laboratory of Computer Architecture, Institute of Computing Technology, Chinese Academy of Sciences, Beijing} \\
\textsuperscript{$\dagger$}\textit{School of Computer and Control Engineering, University of Chinese Academy of Sciences, Beijing}}
}
  
\maketitle

\begin{abstract}
Stream applications are widely deployed on the cloud. While modern distributed streaming systems like Flink and Spark Streaming can schedule and execute them efficiently, streaming dataflows are often dynamically changing, which may cause computation imbalance and backpressure.

We introduce AutoFlow, an automatic, hotspot-aware dynamic load balance system for
streaming dataflows. 
It incorporates a centralized scheduler 
which monitors the load balance in the entire dataflow dynamically and implements state migrations correspondingly.
The scheduler achieves these two tasks using a simple asynchronous distributed control message mechanism and a hotspot-diminishing algorithm.
The timing mechanism supports implicit barriers and a highly efficient state-migration
 without global barriers or pauses to operators.
It also supports a time-window based load-balance measurement and feeds them to the  hotspot-diminishing algorithm without user interference.
We implemented AutoFlow on top of Ray, an actor-based distributed execution framework.
Our evaluation based on various streaming benchmark dataset shows that AutoFlow 
achieves good load-balance and incurs a low latency overhead in highly data-skew workload.
\end{abstract}


\section{Introduction}

Streaming dataflows are widely used ranging from AI applications\cite{moritz2018ray} to website unique
visitors (UV) counting nowadays. These streaming jobs are generally formed as directed acyclic 
graphs (DAGs) by modern streaming systems to be deployed on clusters, each node in the
graph represents a streaming operator defined by users. Streaming framework then
wraps them in each task\cite{carbone2015apache} or actor\cite{moritz2018ray}, 
and initializes network channels, in-memory key-value
store, and other components, and schedules them to physical nodes.

Streaming systems are designed to achieve both low latency and high throughput, while 
latency and throughput can be affected in different ways. An unexpectedly high input source rate that
exceeds the processing capacity of a physical node in the graph will cause latency spikes\cite{kalavri2018three}.
Some physical nodes in the graph may become bottlenecks when there are inappropriate 
settings of parallel instances to operators\cite{hoffmann2018snailtrail}, which will cause backpressure. 
Different fault-tolerance mechanisms may affect the system's throughput and latency in 
different ways\cite{wang2019lineage}. The data-skewness caused by highly uneven access
on a small portion of state
will lead to a hotspot
issue in the streaming job, which often can't be solved well at runtime.

\begin{figure}
  \begin{center}
  \includegraphics[width=0.7\linewidth]{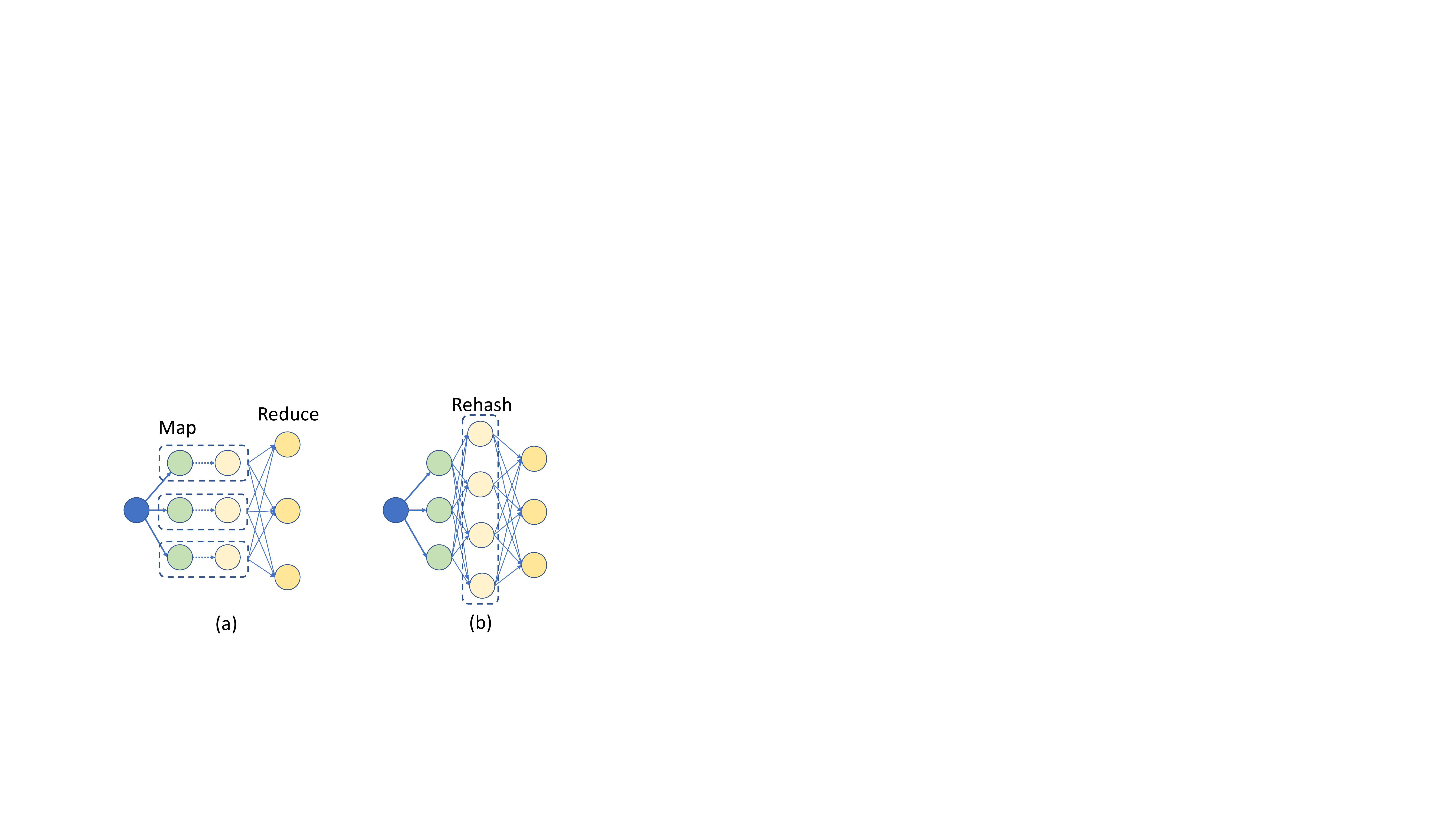} 
  \end{center}
  
  \vspace{-0.4cm}
  \caption{(a) pre-aggregation: insert a local operator that reduce the output of map 
          operator. (b) rehashing: add an extra layer of reduce operators to scatter out
          the potential hotspots}
  \label{img1}
\end{figure}

In this paper, we focus on the data-skewness issue in distributed streaming dataflows. There are two general techniques to handle hotspot
issues statically. One is pre-aggregation (Figure~\ref{img1}), in which the map operators
apply the same reduce logic as the downstream reduce operators on
their portion of data before they send results to downstream.
The pre-aggregation technique reduces pressure to access the state of reduce operator when there
are hotspot issues, while it has obvious drawbacks. Firstly, pre-aggregation requires grouping
a small batch of data, which will cause latency overhead. In some real-time applications, this overhead
may not be acceptable. Secondly, pre-aggregation is only applicable to reduce operations (Count, Sum,
Min, Max), and is not fit for other operations like Join and EventTimeWindow. The other 
technique is rehashing (Figure~\ref{img1}), which adds an extra layer of 
reduce operators to scatter out the potential hotspots 
and merge results in the second layer. The rehashing method has similar drawbacks
to pre-aggregation because it aggregates data twice.

Pre-aggregation and rehashing are static methods that mitigate hotspots issues,
which means the physical graph built at compile time cannot be changed at run time.
While streaming dataflows are long-running jobs in general, and the access pattern
of data to stateful operators is often unpredictable, the static method is inherently 
not the best choice to solve the problem. Flink\cite{carbone2015apache} supports 
rescaling the streaming jobs, 
which we can change the parallelism of the operator instances
and redistribute the state of stateful operators. However, this requires
manual operations such as discovering when and where the hotspot issues raise, 
and how many instances to rescale. And this isn't over yet,
Spark\cite{zaharia2012resilient} 
and Flink\cite{carbone2015apache} have
to halt the whole job graph until the rescaling job finishes, which is the overhead users
have to tolerate. SnailTrail\cite{hoffmann2018snailtrail} adopts Critical Path Analysis 
to successfully detect data-skewness issues
and other computation imbalance problem, while this method has to be integrated 
with a streaming system like Flink or Spark. Chi\cite{mai2018chi} uses 
distributed control messages to coordinate dataflow reconfiguration,
and builds a programmable control plane on top of a streaming system. Although this
is an innovation that firstly integrates control plane and data plane in a system,
when we focus on a specific issue like data-skewness, it's a bit heavy because it
requires explicit barrier on parts of the dataflow graphs when doing reconfiguration.
Megaphone\cite{hoffmann2019megaphone} takes control stream and data stream as inputs 
in each worker simultaneously, and let 
the control inputs to decide which downstream operator to route. When the control 
input modifies the routing table of map operators, it will start the state-migration
between stateful operators. Megaphone supports fluid state migration unlike Flink or Chi,
but its operators require buffering when timestamps of data events are in advance of 
that of the control stream.

We propose
a novel framework AutoFlow to handle the data-skewness problem efficiently.
AutoFlow adopts a similar but lighter control message mechanism.
The timing mechanism supports implicit barriers and a highly efficient state-migration without global barriers or pauses to operators.
It also supports a time-window based load-balance measurement and feeds them to the hotspot-diminishing algorithm without user interference.
These functions are integrated in a centralized scheduler
that monitors the load balance in the entire dataflow dynamically and implements state migrations correspondingly.
 We implemented AutoFlow on top of Ray, an actor-based distributed execution framework. Our evaluation based on various streaming benchmark dataset shows that AutoFlow achieves good load-balance and incurs a low latency overhead in highly data-skew workload. 
Our contributions are the following:
\begin{itemize}
    \item A lightweight centralized timing mechanism that
    generates total ordered control messages and induces implicit barriers.
    \item An efficient state-migration method that only buffers
    minimal data messages.
    \item An effective metric-collection scheme and hotspot-diminishing algorithm to reduce the load-imbalance.
    \item A prototype system AutoFlow integrating the above methods. 
\end{itemize}

The remainder of this paper is organized as follows. Section \ref{sec-2} provides the background. The AutoFlow is described in Section \ref{sec-3}. We present the performance results in Section \ref{sec-4}. Section \ref{sec-5} overviews related work and Section \ref{sec-6}
concludes the paper.

\section{Background and Motivation}
\label{sec-2}

 Modern big data systems can be divided
into two categories, one of them is MapReduce\cite{dean2008mapreduce}, 
Hadoop\cite{white2012hadoop}, and Spark\cite{zaharia2012resilient} that treat
the input data as batches of records, while the other are Flink\cite{carbone2015apache},
Structured Streaming\cite{armbrust2018structured}, Google Dataflow\cite{akidau2015dataflow}
and others\cite{akidau2013millwheel,toshniwal2014storm,kulkarni2015twitter} that treat the data as infinite data streams. A key difference between streaming
and batching is that the streaming systems support event-time operation and watermarks.
Most of the streaming systems focus on how to parallelize the operators to physical
nodes in the clusters and schedule the tasks efficiently. 
This work focuses on hotspot issues, which are often less concerned or well handled in
existing streaming systems.

\subsection{Background} 

Modern streaming systems are more and more deployed on clusters, their goals of 
design first look at the ease of use, performance, scalability and fault tolerance.
Therefore most of their dataflow architectures are often static, that is, they
can act in streaming mode when processing data but act discretely when reconfigurating the 
dataflow graphs. 
Currently streaming systems like Flink can only redistribute 
state by halting the whole dataflow graphs, 
which is a discrete way that incurs latency overhead 
and opposite to  its streaming nature.

Furthermore, existing methods often considers a wide range of areas
such as rescaling, checkpointing and other user-defined functions.
Therefore, current
profiling tools in streaming systems like Flink\cite{carbone2015apache} requires human 
efforts.
Users
may be required to monitor the screen and adjust it when they detect a data-skewness
issue, thus an inevitable latency comes between the detection and repair.

Asynchronous Barrier Snapshot\cite{carbone2015lightweight} is another technique
that has been successfully
applied to Flink\cite{carbone2015apache} as a checkpointing
mechanism. Its core concept has also been used by Chi\cite{mai2018chi} as distributed control
messages to coordinate the reconfiguration stage in stream processing systems.
Distributed control messages act like an asynchronous barrier that coordinate the reconfiguration
process, there may just be a few of the operators are halting at the same time and other
operators work as usual like Chi\cite{mai2018chi}. 
In the case of checkpointing and rescaling, if without the control messages,
we will have to halt all the physical nodes in the dataflow
and resume until the reconfiguration job finishes. The distributed control message mechanism
reduces the overhead by making the control operations asynchronous.



\subsection{Motivation}

For the  data skewness issue in stream processing,
it's desirable to design a lightweight adjusting scheme
that only
requires buffering the related data.  The challenges are how to detect and react 
to data-skewness issues efficiently, and how to migrate state
between straggler and non-straggler without a large overhead. 
The key idea
is to combine data-skewness detecting and reacting as components embedded in the
systems without user interference. 
For detecting the data-skewness issues, we adopt a scheduler to continuously collecting
metrics from stateful operators.
For reacting to the data-skewness issues, we let the scheduler to send distributed 
control messages
to start a state-migration between operators.
These two tasks adopt a control message mechanism in our AutoFlow to determine and adjust the state migration behaviours
between the stragglers and non-stragglers efficiently.


\section{AutoFlow design}
\label{sec-3}
\begin{figure}[b]
  \begin{center}
  \includegraphics[width=1.0\linewidth]{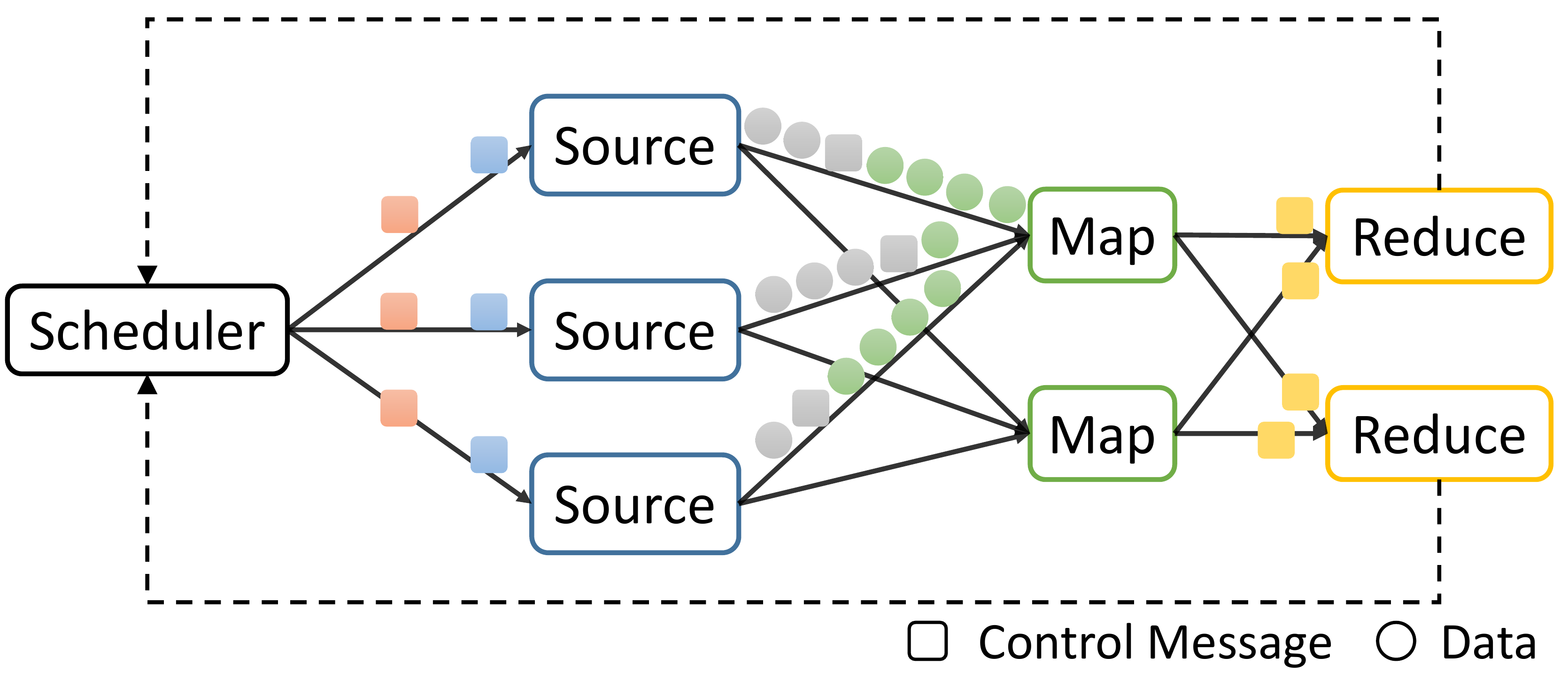} 
  \end{center}
  
  \vspace{-0.4cm}
  \caption{ AutoFlow model. The scheduler is embedded as an operator in the
          dataflow graph, and continuously broadcasts control messages with 
          increasing timestamps to source operators. The source operators
          flush data events until their timestamps exceeds the control
          event, and then propagate the control message indicated a data 
          boundary.}\label{img3}
\end{figure}


\subsection{Overview}
\label{overview}

AutoFlow's dataflow model is similar to most of the streaming
systems like Flink\cite{carbone2015apache}, 
Spark Streaming\cite{zaharia2013discretized} and
Google Dataflow\cite{akidau2015dataflow}. It consists of two categories
of operators. The first one is stateless operators such as map, flatmap, and filter, which 
does not hold any local mutable state. The second is stateful operators such as join,
reduce, time-window. The stateful operator is one of the core concepts in 
streaming systems because
many functionalities like checkpointing, state management is built for it.

The AutoFlow model is depicted in Figure~\ref{img3}.
It abstracts the dataflow model as a DAG graph.
The major part is a centralized lightweight scheduler that is
embedded in the dataflow graph as a new operator.
The scheduler is connected to all source operators
 and all stateful operators.
The scheduler carries out two tasks:
dynamically monitor the load balance in the entire dataflow and implement state migrations correspondingly.
To achieve these goals, the scheduler incorporates
a simple timing mechanism and a hotspot diminishing algorithm.

\subsection{Timing Mechanism}
\label{timing}

The scheduler performs as a timer.
It periodically generates empty control messages 
(the square items Figure~\ref{img3})
that are sent to all downstream operators,
in particular, all source operators.
Each message is attached with a timestamp.
After receiving a control message, the source operator   
inserts it in its dataflow (circle items).
The messages spread over the entire dataflow graph in a similar way
and finally, return back to the scheduler.

The fields
of the control messages are listed in Table~\ref{table1} which are all immutable. If the scheduler operator sends an empty control message, the fields of migration, send, 
receiver and slot ids in the message are all empty, but the rest of the fields are
required because it indicates a boundary of a period of datastream for further use.
The field `migration' indicates whether we
perform state-migration or not.

The function of these messages is three-fold.
First, previous schemes often utilize partially ordered timestamps.
On the contrary, control messages are launched by a centralized scheduler
in AutoFlow and totally ordered.
This feature serve as a base of the following functions.

Second, control messages indicate implicit barriers.
Control 
messages come from all upstream operators that have an equal timestamp 
identify an implicit control barrier.
Specifically, when the downstream operator reducer-0 recevie
a control message from one of its input channels that has the content 
(\textit{event\_time}=200, \textit{sender}=reducer-0, \textit{receiver}=reducer-1,
\textit{slot\_id}=2).
The control message implies that: (1) there will be no data events with timestamps
\textit{t}$<$200 coming from this channel, and reducer-0 can safely migrate the state of
the 2nd slot to reducer-1 after it receives the same control messages from all its
input channel. (2) the data events with timestamps \textit{t}$>$200 will be routed
according to the update routing table in the map operators, the reducer-0 doesn't have to worry if
it will receive coming data events that are routed to the migrated slot it has sent.
These two implications are critical to designing an efficient state-migration scheme.  
The detailed description is presented in Section~\ref{state-migration}.

Third, a control message associated with a timestamp indicates a boundary of data records in a period of time.
The source operator simply injects the control message
in the data message flow and send it to all downstream operators.
It is equal to group the data messages between two continuous control messages
to an atomic data set that is processed identically.
Since a control barrier message will be eventually broadcasted to all stateful operators,
this property provides an implication for measuring the load-imbalance.
Specifically, 
each stateful operator keeps collecting statistics data such as processing counts of each key slot. 
When a stateful operator receives all the empty control messages that carry the same timestamp 
from upstream, it will attach statistics results to the control message and send it back to the scheduler for further analysis.
The scheduler keeps collecting metrics
from stateful operators,
analyzes the collected profiling data and 
employs an algorithm to optimize the load-balancing dynamically.
It attaches a state-migration instruction to the next control message when detecting a hotspot issue. 
The detailed description is presented in Section~\ref{hotspot-algorithm}.




Another significant difference between Autoflow and other systems is 
that we do not necessitate a timestamp associated with the data message.
The reason is that Autoflow only models a DAG graph without loops
and assumes a FIFO communication channel.
However, we can extend the timing scheme and support
complex scenarios like dataflow graphs with loops.
Specifically, we can utilize a simple logical timestamp scheme which is coordinated between
the control messages and data messages.
When a source operator receives an empty control message from the scheduler, it will
continuously send data events that have smaller timestamps than of the control message to 
downstream operators. When the data left in the source have timestamps that are larger than
that of the control message, the source operator will propagate the control message to downstream
and fetch another control message that has a larger timestamp.
Other operators can be augmented in a similar way.
We leave this extension as future work.


%
%

\begin{table}
  \renewcommand\arraystretch{0.75}
\small
\centering
\begin{tabular}{ccc}
  \toprule  
  Field& Type& Description\\
  \midrule  
  $event\_type$& string&  data record or control event\\
  $evnet\_time$& long&  the logical timestamp\\
  $migration$& bool&   trigger state-migration or not\\
  $sender$& string&  the sender name of the state\\
  $receiver$& string&  the receiver name of the state\\
  $slot\_ids$& list of strings&  which state slots to send\\
  \bottomrule 
 \end{tabular}
 \label{table1}
 \caption{Control messages sent by the scheduler operator}
\end{table}

\begin{figure}[b]
  \begin{center}
  \includegraphics[width=1.0\linewidth]{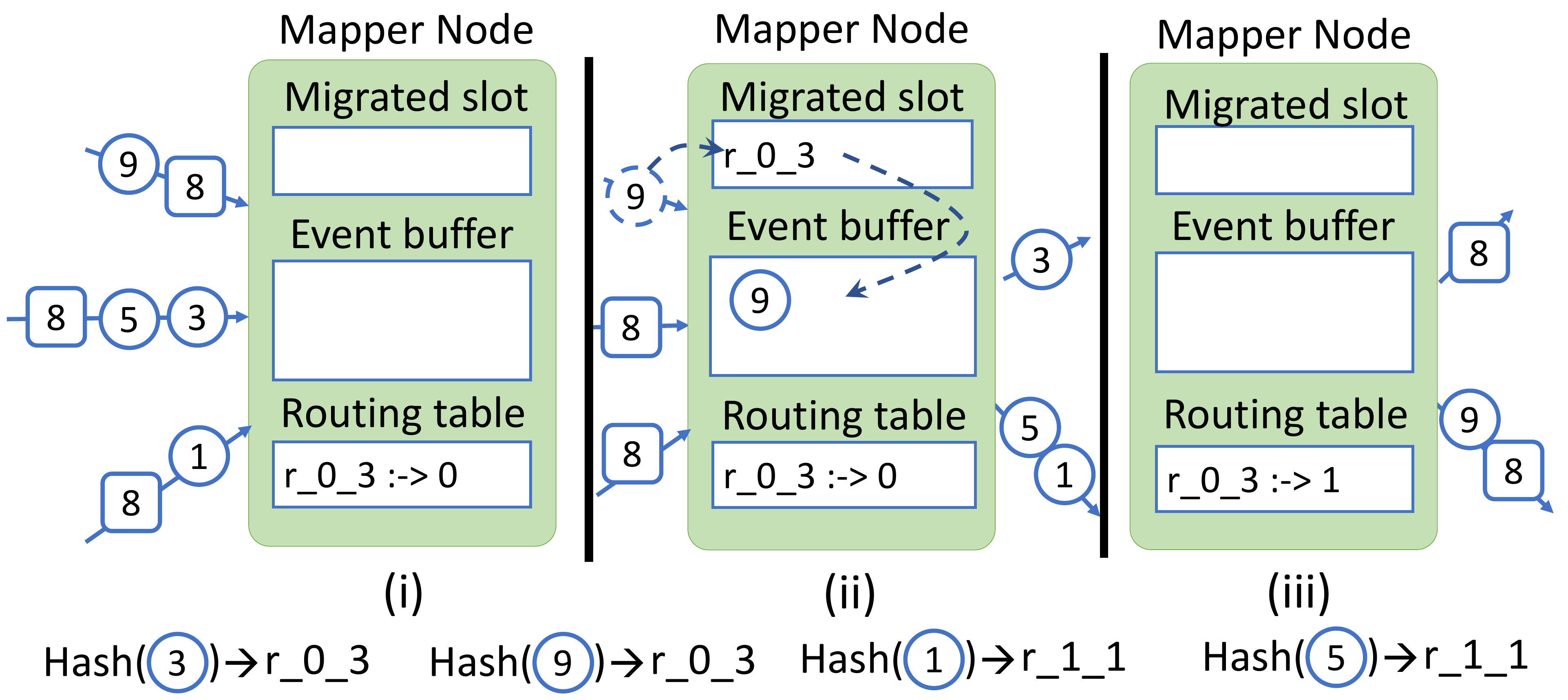} 
  \end{center}
  
  \vspace{-0.4cm}
  \caption{ Processing procedure of control operations in the map operator. 
  (i) map operator receives a control message from uppermost channel
  (ii) mapper buffers the data events following the control message
  (iii) mapper finishes processing control messages and broadcast it to downstream.}\label{img4}
\end{figure}

\begin{algorithm}[htb]
  \caption{ Control Operation in Map Operator }
  \label{algo:1}
  \begin{algorithmic}[1]
    \Function {process}{event}
      \If {$event["event\_type"]$ is $"ControlMsg"$}
        \State event\_time = event[$"event\_time"$]
        \If {marker.get(event\_time) is $None$}
          \State marker[event\_time] = 1
            \If {event[$"migration"$] is not $None$}
               \State Init(migrated\_slot, event\_buffer)
            \EndIf
        \Else
          \State marker[event\_time] += 1
        \EndIf

        \If {marker[event\_time] == num\_input}
          \If {event[$"migration"$] is not $None$}
            \State slot\_ids = event[slot\_ids]
            \For {slot\_id in slot\_ids}
              \State receiver = event[$"receiver"$]
              \State routing\_table.set(slot\_id, receiver)
              \State flush(slot\_id, receiver)
            \EndFor
          \EndIf
          \State broadcast\_control\_msg(event)
        \EndIf
      \ElsIf {event[$"event\_type"$] == $"DataMsg"$}
        \State slot\_id = event[$"slot\_id"$]
        \State event\_time = event[$"event\_time"$]
        \If {slot\_id is in migrated\_slot \&\& event\_time $>$ marker[event\_time]}
          \State self.event\_buffer.append(event)
        \Else
          \State process\_data(event)
        \EndIf
      \EndIf
    \EndFunction
    
  \end{algorithmic}
\end{algorithm}

\subsection{State Migration}
\label{state-migration}

We now present an efficient state-migration implementation
with the assistance of the timing mechanism.
When an operator receives a control message
with a state-migration instruction from one of its input channels, 
it can expect the same control messages from all other channels.
With the implicit barrier induced by the timing mechanism,
it just needs to buffer subsequent data messages which are
routed to the migrated slot indicated by the state-migration instruction and from the same channel.
Other data messages can be immediately processed as usual.
Moreover, we can process multiple
state-migration control messages at the same time.
Autoflow does not need to 
 stop the data input channel or require global barriers.
The implementation details for
 stateless and stateful operators are discussed in the following two parts respectively.


\subsubsection{Stateless operator}
\label{map}
Although stateless operators like map, filter do not hold any state of the dataflow, 
it's their job to decide which downstream stateful operator to send when they
processed a data record. There are often many parallel instances of operators in a streaming 
job for horizontal scalability, so it's a common way that each map operator has
the same immutable hashing function that acts as a static routing table shared across 
a cluster of workers.  
We split the distribution of state into many slots like the virtual nodes in
consistency-hashing and each slot has a unique id.
The slot is the atomic unit in our state-migration process. We only migrate slots between 
stateful operators when doing state-migration.

Algorithm~\ref{algo:1} shows how to process control messages
in map operators.  When the "migration" field is None, we just record
it in the marker and continue to process data, and when the map operator receives all 
the same control messages from its upstream channels (L15), it will broadcast it to all
its downstream operators. Thus, if there's no migration event happens, the map operator
does nothing different from when there's no control message mechanism.

Algorithm~\ref{algo:1} registers the migrated slots (L6) when there's a state-migration event comes, It then buffers records from channels that have already received the control message
if the corresponding slot will be migrated (L27).
Finally, when the map operator receives  control messages from all its input channels,
it means an implicit barrier, and all data events in the buffer should be routed with respect to the updated routing table. 
So it is safe to change the routed table, broadcast the
control messages to all its downstream reducers, and flush
all records in the event buffer (L13-18).
In summary, we only have to buffer data events that
are routed to the migrated slot temporarily. 

We demonstrate an example shown in Figure~\ref{img4} where the
map operator receives state-migration events interleaved with other data events
from three input channels.
 The slot named
"r\_x\_y" means the y-th slot of reducer-x. The bottom of the figure is the hashing results of all
data events and the routing table tells the slots owned by which reducer instance.
In Figure~\ref{img4} (i), 
when the map operator receives a square control message with timestamp "8" from one 
uppermost channel, it parses the message and gets the name of the migrated slot "r\_0\_3", which 
will be migrated from reducer-0 to reducer-1. Then the map operator writes down the slot id on
its migrated slot list to indicate which data events need to be buffered. The map operator
buffers the circle data event from the uppermost channel in Figure~\ref{img4} (ii)
according to the migration instruction.
The data event hashed to the slot "r\_0\_3" in the middle input channel
is sent to reducer-0 as usual since the map operator has not received the control.
Figure~\ref{img4} (iii) illustrates that after
the map operator receives the control messages with timestamp "8" from all its input channels.
it changes the routed table, broadcast the
control messages and flush
all records in the event buffer to reducer-1.

\begin{algorithm}[htb]
  \caption{ Process Operation in Stateful Operator }
  \label{algo:2}
  \begin{algorithmic}[1]
    \Function {process}{event}
      \If {$event["event\_type"]$ is $"ControlMsg"$}
        \State event\_time = event[$"event\_time"$]
        \If {marker.get(event\_time) is $None$}
          \State marker[event\_time] = 1
            \If {event[$"migration"$] is not None}
              \If {event[$"receiver"$] == my\_id}
                \State Init(migrated\_slot, event\_buffer)
              \EndIf
            \EndIf
        \Else
          \State marker[event\_time] += 1
        \EndIf

        \If {marker[event\_time] == num\_input}
          \If {event[$"migration"$] is not $None$}
            \If {event[$"sender"$] == my\_id}
              \State send\_migration(event)
            \ElsIf {event[$"receiver"$] == my\_id}
              \State async\_recv(event[$"sender"$])
            \EndIf
          \EndIf
          \State broadcast\_control\_msg(event)
        \EndIf
      \ElsIf {event[$"event\_type"$] == $"DataMsg"$}
        \State slot\_id = event[$"slot\_id"$]
        \If {slot\_id is in migrated\_slot}
          \State self.event\_buffer.append(event)
        \Else
          \State process\_data(event)
        \EndIf
      \EndIf
    \EndFunction
    
  \end{algorithmic}
\end{algorithm}

\begin{figure*}
  \begin{center}
  \includegraphics[width=1\linewidth]{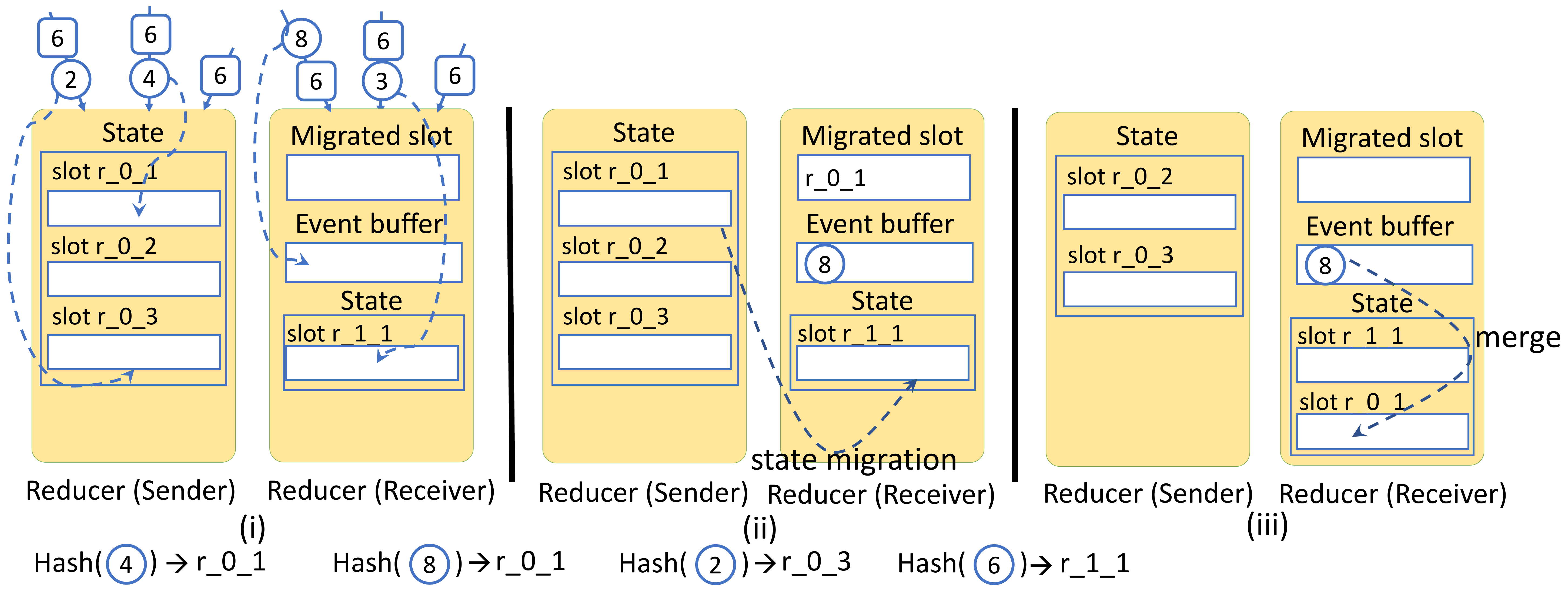} 
  \end{center}
  
  \vspace{-0.6cm}
  \caption{ Processing procedure of control operations in reduce operator. 
  (i) Reduce operators receive data events interleaved with control messages
  (ii) Reduce operator(right) write downs slot\-"r\_0\_3" in the migrated slot 
  and initializes
      the event buffer when processing the control event. In the mean time reduce
      operator(left) triggers a state-migration to the right.
  (iii) After reduce operator(right) receives state from the left, it then merge
    the events in the event buffer.}
  \label{img5}
\end{figure*}

\subsubsection{Stateful operator}
\label{reduce}
Many core concepts in streaming systems like state-management, flow-control, and
checkpointing are highly related to stateful operators that hold local mutable
state in a key-value store. Unlike stateless operator where we barely encounter a hotspot
issue because we can easily adapt schemes like random shuffle, round-robin and 
rebalance between operators to ensure load-balance, stateful operators
often face the data-skewness problem that requires state-migration.


Cooperated with the stateless operator, specifically Algorithm \ref{algo:1},
we know that (1)
if a stateful operator has received a migrated control message, it will not receive 
the following event that routed to the slot. Our mechanism
ensures the upstream stateless operators will route the event to the new operator.
(2) the stateful operator will trigger a state-migration process after it receives all the control messages. With these two features, we present a similar algorithm for stateful operators (Algorithm~\ref{algo:2}) and describe it
with an example in Figure~\ref{img5} .




In Figure~\ref{img5} (i) the left reducer firstly receives
a control event "6" from the uppermost channel and record it in the marker (L4).
Although the data event "4" from the middle channel is routed to the migrated slot is "r\_0\_1", 
the reducer has not receive the control event from it yet, so it will process the event as usual 
and change the state in "r\_0\_1". In the meantime, the right reducer receives the same control
event and initializes buffer and migrated-slot list (L6-12), when there comes a data event "8" right after
the control event that routed to slot "r\_0\_1", the reducer just buffers it. 
In Figure~\ref{img5} (ii) when the left reducer receives all control events, it will starts migration
and send slot data "r\_0\_1" to the received reducer (L19-25). In Figure~\ref{img5} when the right reducer
receives the slot (L27), it will merge the data events in the buffer to the slot. It's worth mentioning 
that the migration process is asynchronous which means that the reducers are processing other data 
events concurrently.

\subsection{Hotspot-diminishing}
\label{hotspot-algorithm}
An external controller that widely used in big data systems is for task
scheduling, resource management, and monitoring runtime status of the cluster. 
It's a centralized way that can coordinates the tasks in a cluster directly.
In AutoFlow we use scheduler as an operator in a more fine-grained way to 
tackle the data-skewness issue. Specifically, in the scheduler operator, we integrate
monitoring of the runtime status in stateful operators with a feedback control 
component that reacts to the dataflow efficiently.

There are lots of ways to monitor a streaming system such as collecting the 
throughput and latency, monitoring the queue size of the input channels, and probing the 
data input or output rates of operators. Some of them either do not reflect the situation
accurately\cite{hoffmann2018snailtrail} or are not efficient due to the natures of 
distributed systems such as network latency.
As the control messages act like watermarks, we let the stateful operators send metrics
such as the processing counts of each slot, and the total processing count to the scheduler operator
when they reach an
implicit barrier triggered by the empty control messages.

Common monitoring systems collect metrics such as memory consumption in each task,
throughput and latency per record in each operator, size of the network buffers 
and so on. These collected metrics are often shown in the dashboard for users.
When we build an integrated monitoring and feedback control component tackling the
data-skewness, there're three questions we need to consider:


\textbf{What information should we get?} When there's a data-skewness issue, 
we will know that
a large amount of data requests are routing to a few of the stateful operators.
Throughput and latency are the two common metrics in streaming systems. 
Throughput represents the data output rates of an operator, 
which will reach a nearly constant when the data input rate 
exceeds or just reached its processing capacity. That is, 
we can hardly tell the difference between
stragglers and non-stragglers just according to their throughputs. The latency of the
stragglers will grow high beyond the non-stragglers. But these metrics are not
enough for a feedback algorithm to make decisions of state-migration. In AutoFlow 
we collect the number of processing records of each slot 
in a period of time from every stateful operator.

\textbf{How do we get?} 
A stateful operator sends metrics with the timestamps each time when it
has received control messages from all channels. In the scheduler operator, 
we start a thread that keeps
listening from stateful operators asynchronously. When the scheduler 
receives metrics messages from all stateful operators, it indicates a boundary of time.
It then feeds those metrics to our hotspot-diminishing algorithm.

\textbf{How to analyze the metrics?} 
As discussed above, the scheduler operator receives
the processing count of each slot in every interval. 
While the dataflows are often changing dynamically,
The current metrics data it receives might not always predict the situation in the 
next second correctly. For example, in a flash-sale activity online there might 
just be a latency spike in a task and vanished in a few seconds, while in other 
case the stragglers might last for a long time. 
Setting a better window-length parameter can help to predict the situation more accurately.
For a wider range of application, 
we adopt a time-window based feedback algorithm in AutoFlow.
In a time-window scheme, the algorithm can combine the information in a long time
or in a short period of time.


\begin{algorithm}[htb]
  \caption{ Main Part of the Control Algorithm }
  \label{algo:3}
  \begin{algorithmic}[1]
    \Require
      $total\_count:$ $\{"worker\_id": count\}$ indicates processing
      count of each stateful operator respectively

      $slot\_count:$ $\{"worker\_id": \{"slot\_id": count\}\}$
    \Ensure
      $migrate\_slot:$ a list of slot\_ids that will be migrated

    \Function {process\_window}{total\_count, slot\_count}
      \State max\_id = max($total\_count$)
      \State min\_id = min($total\_count$)
      \State migrated\_slot = []
      \State diff = total\_count$[$max\_id$]$ - total\_count$[$min\_id$]$
      \If {diff / total\_count[max\_id] $>=$ factor}
        \State gap = diff / 2
        \While {gap $>$ 0 \&\& slot\_count is not None}
          \State slot\_id = max(slot\_count[max\_id])
          \State num = slot\_count[max\_id][slot\_id]
          \If {gap $>=$ num}
          \State migrated\_slot.append(slot\_id)
          \State gap = gap - num
          \EndIf
          \State slot\_count$[$max\_id$]$.pop(slot\_id)
        \EndWhile
      \EndIf
      \State \Return migrated\_slot
    \EndFunction
    \end{algorithmic}
\end{algorithm}


\textbf{How to react?} 
For static workloads, it is often satisfied to shuffle the state between
all workers with a global barrier to reach a perfectly load-balance.
However, this scheme will incur a large overhead for long-running and  dynamically changing dataflows.
In Autoflow, we tend to act dynamically and continuously that we do not need to migrate 
a large amount of state. 

Our hotspot-diminishing algorithm is presented in Algorithm~\ref{algo:3}. 
It only adjusts the workloads between the operators with the maximum
and minimum metric, in particular, the total number of processing records in a time-window. 
Specifically, when the gap between the max-operator and 
the min-operator exceeds a predefined value (L6),
we utilize the first-fit heuristic (L8-14) of the knapsack problem to 
pack slots whose total size approximately equals to half of the gap.
Finally, the scheduler operator will serialize the migration meta-info
in the control event and propagate it in the dataflow.
Although we illustrate one algorithm implemented in AutoFlow, there are also
many other load-balance schemes that can be adopted in our framework. Introducing
user-defined function as a load-balance API is one of its future work.

\subsection{Implementation}
AutoFlow is built on top of Ray\cite{moritz2018ray}---an actor-based distributed 
execution framework. We implemented the core logic of operators and transform
the logical dataflow graph to a physical graph through Ray's Python bindings, in which
each operator is wrapped in an actor. Each actor is connecting with each other through
grpc\cite{grpc} in Ray, when an operator sends a message to its
downstream channel, a \textit{remote} function will be called through the backend 
of Ray. Each operator in the logic graph 
of dataflow will be deployed in a cluster combined with a Python process.
Currently, we implemented the (de)serialization of data events, 
and the state of stateful operators in Python's standard library.
Moving them to a more efficient
programming language's backend and adopting a high-performance key-value store are 
in our future work.
In AutoFlow we bypassed the scheduling process of Ray's backend
by turning on the \textit{direct\_call} mode in the \textit{remote} function's API.
In \textit{direct\_call} mode, Each \textit{remote} function call is a 
Remote Procedure Call (RPC) to another process. At the start of the deployment, Ray
will connect each actor through gRPC\cite{grpc}.

\begin{figure*}
  \begin{center}
  \includegraphics[width=1.0\textwidth]{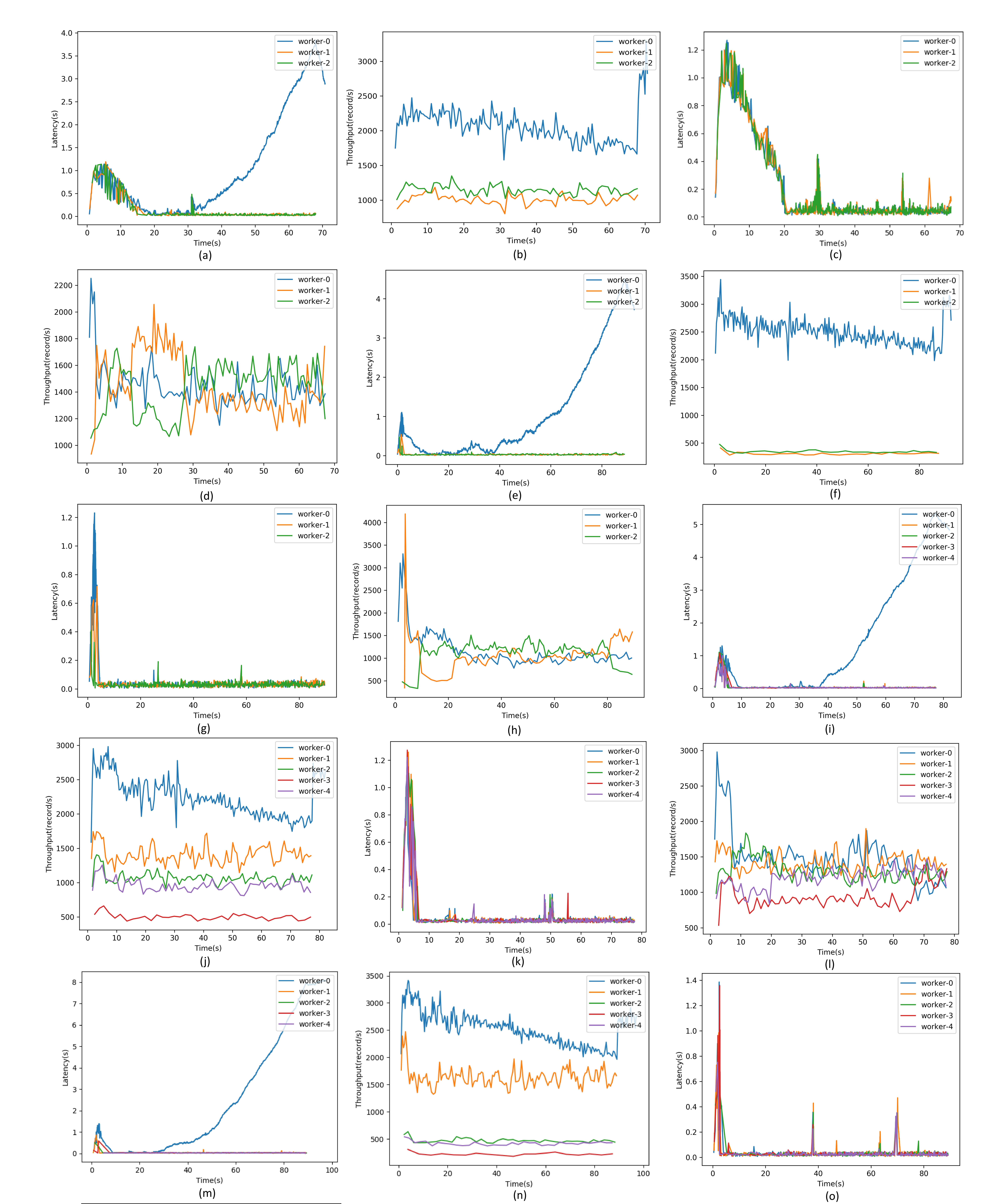} 
  \end{center}
  
  \vspace{-0.5cm}
  \caption{(a)(b) 50\% skewness in 3 reducers without dynamic load-balance. 
           (c)(d) 50\% skewness in 3 reducers with dynamic load-balance.
           (e)(f) 80\% skewness in 3 reducers without dynamic load-balance.
           (g)(h) 80\% skewness in 3 reducers with dynamic load-balance.
           (i)(j) 60\% skewness in 5 reducers without dynamic load-balance.
           (k)(l) 60\% skewness in 5 reducers with dynamic load-balance.
           (m)(n) 80\% skewness in 5 reducers without dynamic load-balance.
           (o) 80\% skewness in 5 reducers with dynamic load-balance.}
  \label{img9}
\end{figure*}

\section{Evaluation}
\label{sec-4}
We evaluated the AutoFlow model on the generated Nexmark benchmark dataset.
To measure the benefits of AutoFlow's dynamic load-balance scheme, our evaluation
falls into four categories. First, we ran a stateful query on various workloads with
fixed skewness percentages to show the AutoFlow can achieve load-balance through
our control message mechanism (Section~\ref{static}). Second, we tested our model on
workloads with a suddenly changed extent of skewness to show the reaction speed
of our algorithm (Section~\ref{spiking}). Third, To an unpredictable, 
rapidly changing workload, our algorithm can handle it to what extent? (Section~\ref{dynamic})

\subsection{Experimental setup}
We run all our experiments on up to three machines on the Aliyun ECS cluster, each 
is an m5.2xlarge instance with 8 vCPUs and 32GB of RAM, running on Ubuntu 18.04.
The schema in Nexmark benchmark is a model that consists of three 
entites: \textit{Person, Auction, Bid}, which simulates a public auction. We generated
the dataset according to the opened source code from\cite{nexmark}. There many standard
queries in Nexmark benchmark, such as a simple map (\textbf{Q1}), a simple filter (\textbf{Q2}),
an incremental join (\textbf{Q3}). We chose to test a source-map-reduce model on our workloads
as the following stateful query:
\begin{verbatim}
  stream = env.source(record)
              .map(record => bids)
              .keyby(bids.id)
              .reduce(state[bids.id] 
                      += bids.price)
\end{verbatim}
And each operator has the same parallel instances in our settings.

\subsection{Fixed skewness workload}
\label{static}
We tested various workloads with different skewness percentages on AutoFlow,       
and compared the results between turning off and turning on the load-balance 
algorithms. The latency and throughput in each workload are shown in 
Figure~\ref{img9}. The latency is sampled once every 100 records, and the 
throughput is recorded once every 1000 records. 

Figure~\ref{img9} (a) shows an 
increasingly high latency overhead when 50\%
of data events were routed to one of the reducers, and Figure~\ref{img9} (b) shows
the throughput in the hotspot worker is obviously higher than the others, the 
load-balance algorithm was not used in this workload. When we turned on the 
load-balance scheme in the scheduler operator, the data-skewness issue is 
alleviated according to Figure~\ref{img9} (d). We detected a slightly latency
spike and a throughput dropping down of reducer-1 (\textit{t}=30s in Figure~\ref{img9}(c)(d)), 
this indicated state-migration between
reducers has happened. Figure~\ref{img9} (e) and Figure~\ref{img9} (f) illustrate a high skewness workload 
that most of the data events were routed to one reducer, and the problem was
solved well from the start. Although the migration overhead is hidden at the start
of Figure~\ref{img9} (g), we detected a slight latency overhead of state-migration
happened at \textit{t}=25 (Figure~\ref{img9}(h)).

The workloads depicted in Figure~\ref{img9} (i)-(o) is running on 5 workers of each
operator. The x\%-skewness in those workloads means that x\% of the data events
are routed two of the reducers. Figure~\ref{img9} (i) depicts the overhead caused
in the 60\% skewness workload, and the problem was solved through migration with
some slight latency overhead (\textit{t}=50 in Figrue~\ref{img9}(k)). The rest
of the figures also show the alleviation of data skewness in AutoFlow.

In the above workloads, we illustrate that the AutoFlow model reaches better a load-balance.
while in the real-world applications the dataflows are often dynamically changing,
we need to generalize our algorithm to fit in the unpredictable environment.

\subsection{Spikingly changing workloads}
\label{spiking}
\begin{figure}
  \begin{center}
  \includegraphics[width=1.0\linewidth]{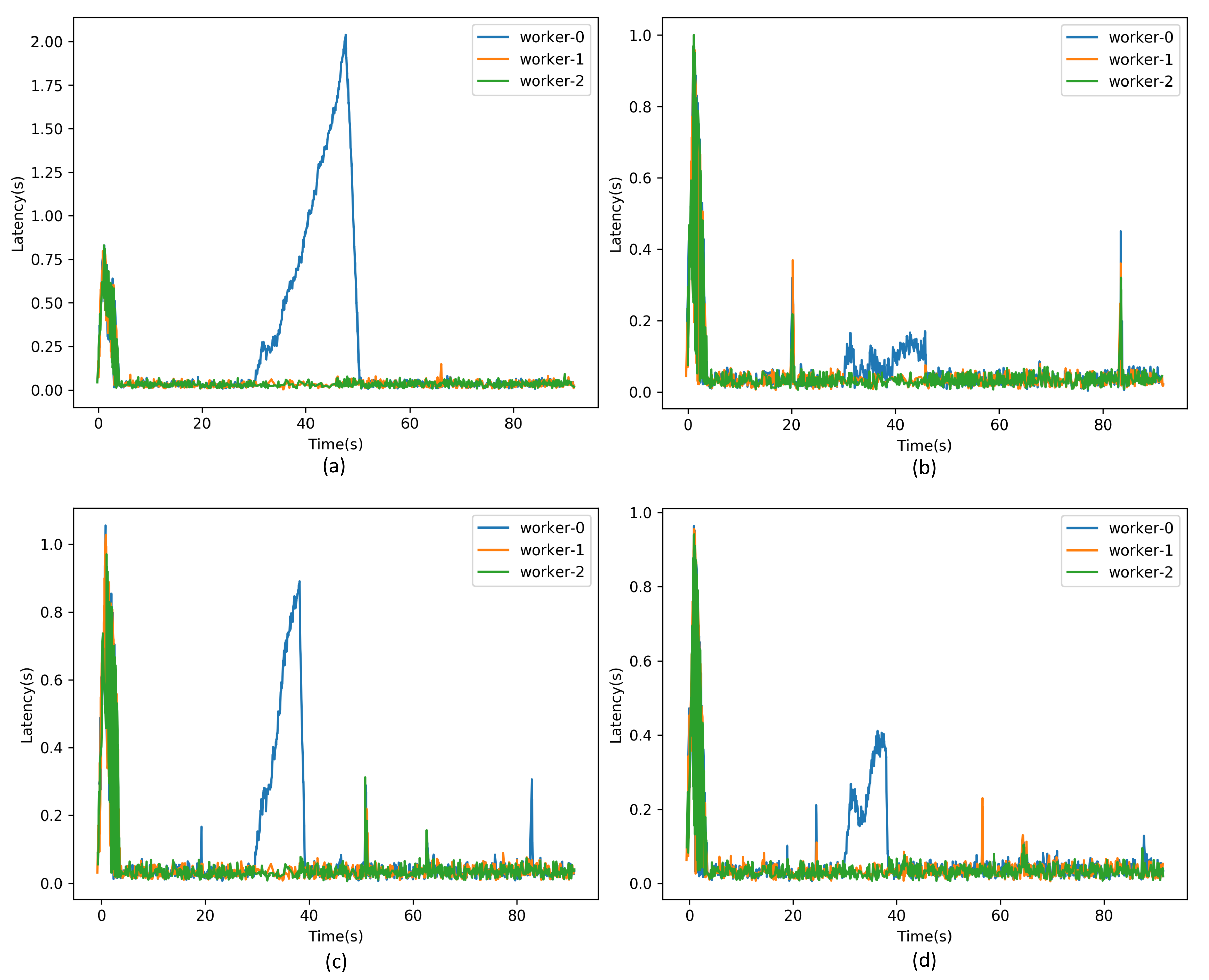}
  \end{center}
  
  \vspace{-0.5cm}
  \caption{(a) A spiking event occured and lasted for 20 seconds.
           (b) the AutoFlow reduced the latency overhead caused
                by the spiking event with a small \textit{window-length} parameter.
           (c) a spiking event occured and lasted for 10 seconds occured.
           (d) the AutoFlow relaxed the issue with a smaller \textit{window-length}
                parameter than that of (b).
           }
  \label{img10}
\end{figure}
In real-world scenarios, we often encounter spiking events in which a large number
of requests are flushed to servers for just a few seconds. These also cause
hotspot issues in most of the time. Solving them requires a faster reactive
method to detect the issues from the beginning.

The feedback control algorithm proposed in Section~\ref{hotspot-algorithm} has a \textit{window-length}
parameter exposed to users that can generalize to these spiking events. The 
\textit{window-length} parameter indicates the duration the scheduler operator
collects metrics for. When the dataflow is stable or slowly changing, we can
collect the metrics from a broad range of time by setting the \textit{window-length}
parameter bigger, and the scheduler operator can make decisions better
according to those metrics. However, when the dataflow has spiking events, the effective
metrics are concentrated on a short range of time. In this case, it's wise to 
set the \textit{window-length} smaller to better suited to the fast-changing workloads.

As depicted
in Figure~\ref{img10}, we showed workloads that have spiking events lasted for
20 seconds (Figure~\ref{img10}(a)) and 10 seconds (Figure~\ref{img10}(c)). In 
Figure~\ref{img10} (b), we set a smaller \textit{window-length} and got better
results compared to Figure~\ref{img10} (a). The latency overhead is greatly reduced
by our feedback control algorithm. However, to the workload with shorter reaction
time in Figure~\ref{img10} (c),
our algorithm can only relax it to some extent. There are some reasons for that:
(1) our algorithm is not efficient enough for this workload, (2) there's a delay
from sending and receiving metrics, and a delay from sending control messages
and performing state-migration.
Building a more reactive scheme is in our future work.

\subsection{Dynamically changing workloads}
\label{dynamic}
\begin{figure}
  \begin{center}
  \includegraphics[width=1.0\linewidth]{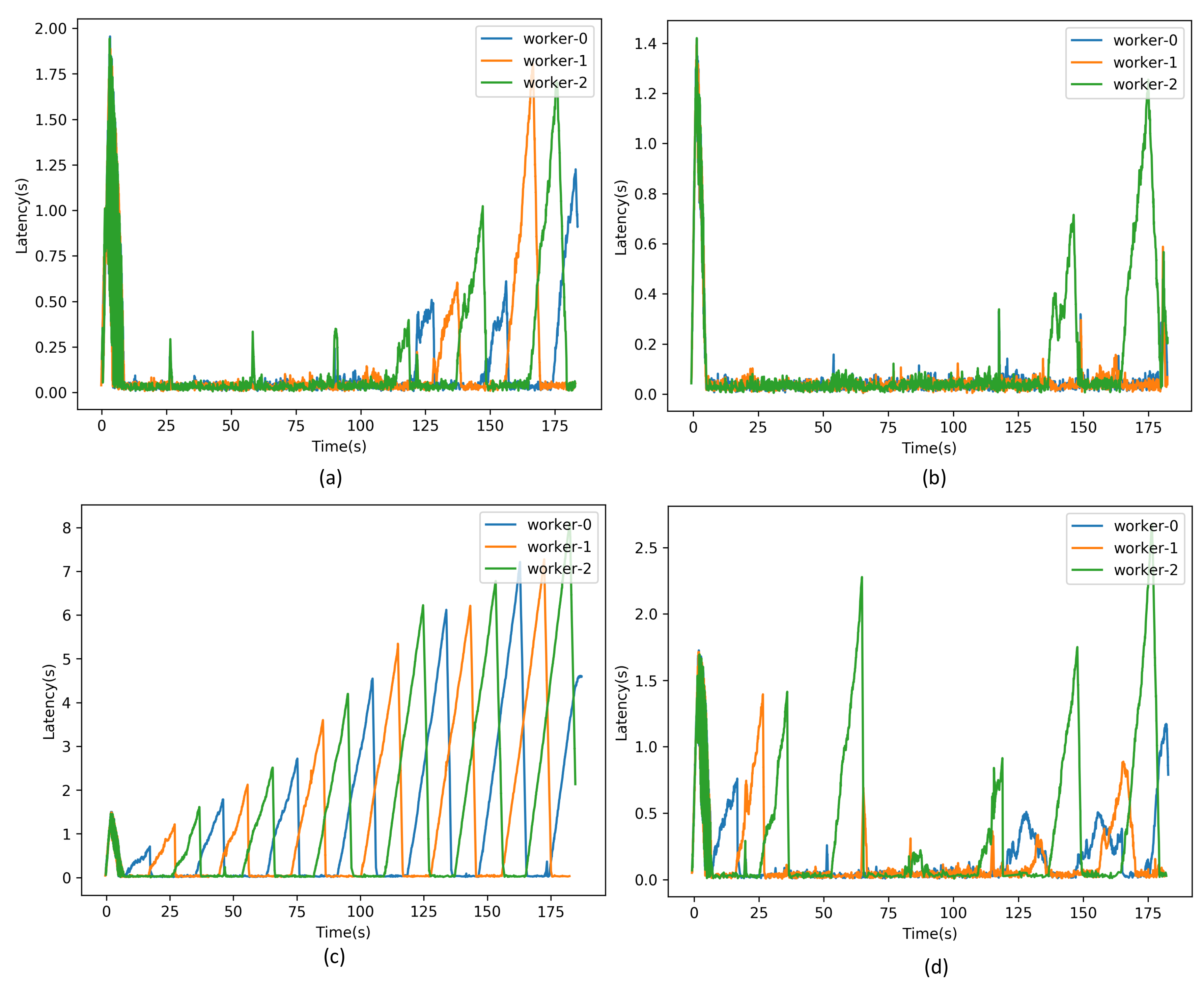}
  \end{center}
  
  \vspace{-0.5cm}
  \caption{(a) 50\% skewness dynamically changing workload.
           (b) the AutoFlow reduced the overhead to some extent.
           (c) 80\%, round-robin skewness workload. Each reducer took turns
                being a hotspot.
           (d) the AutoFlow fixed some hotspots but some other still existed.
           }
  \label{img11}
\end{figure}

Streaming dataflows are long-running jobs and they often change dynamically. To
see if our algorithm can fit into the dynamically changing environment, we generated
the workloads that have hotspots in every moment and the location of the hotspot changed
dynamically. 
At any time, there was a reducer that becomes the hotspot in the dataflow depicted in Figure~\ref{img11} (a)
and Figure~\ref{img11} (c), It's worth mentioning that the situation in Figure~\ref{img11} (c)
was far worse than that in Figure~\ref{img11} (a), because the time when the latency
overhead rises depends on both the skewness percentage and the source input rate. Figure~\ref{img11} (b)
shows fixed the hotspot issues but left for one, this may due to the efficiency of our algorithm.
In Figure~\ref{img11} (d), AutoFlow detected some of the hotspots and reduced
the overhead greatly compared to Figure~\ref{img11} (c), but some other hotspots still existed.
Imagine a case, a straggler migrates state to a non-straggler for load-balance, after a while,
the non-straggler becomes the hotspot, and the situation may become worse.

Although we might not meet the above extremely workloads in real-world applications, we
shows the generality of our algorithm to some extent.

\section{Related work}
\label{sec-5}
There are two types of work on building more reacting, more efficient
streaming systems. One is building a monitoring system or algorithms that can detect issues
or failures in the system faster and more accurately
\cite{kalavri2018three,hoffmann2018snailtrail,toshniwal2014storm,dai2011hitune,garduno2012theia}. 
The other is finding a more 
efficient way of doing specific things like checkpointing, 
rescaling\cite{carbone2015lightweight,floratou2017dhalion,zaharia2013discretized,castro2013integrating}.

\textbf{Reconfiguration.}
Flink\cite{carbone2015apache} currently only supports rescaling when the 
whole dataflow graph is stopped. Megaphone\cite{hoffmann2019megaphone}
employ a similar fluid-
migration scheme
but it  lacks of a controller
and  needs to block more the data message, which also incurs overhead.
SEEP\cite{castro2013integrating} integrated a checkpointing
mechanism that can act asynchronously with the rescaling operation to reduce the overhead.
Dhalion\cite{floratou2017dhalion} provided a control policy mechanism that allows users
to define their own policy and self-tunes a streaming job with their needs. Chi\cite{mai2018chi}
integrated control messages with data messages on a programmable control plane
in distributed stream processing. However, our AutoFlow differs from the
above methods in: 1.we focus on tackling the data-skewness issue through distributed control
message mechanism, we use control message only for state-migration but the other
control operations. 2.our AutoFlow only requires buffering the migrated part
of the data events when doing the migration, other methods either require halting 
the whole dataflow graph or blocking channels of some operators.

\textbf{Monitoring.}
DS2\cite{kalavri2018three} provided an external controller that supports both
automatic scaling and monitoring, SnailTrail\cite{hoffmann2018snailtrail} adopted
Critical Path Analysis (CPA) to analyze the bottleneck of the streaming dataflows.
AutoFlow embedded a lightweight scheduler as an operator to 
continuously analyze hotspots in the dataflows and sends control messages to 
perform state-migration between stragglers and non-stragglers. Thus, our model is an
integrated approach to tackling a specific problem.

\section{Conclusion}
\label{sec-6}
In this paper, we proposed and evaluated AutoFlow, a hotspot-aware dataflow
model that support dynamic load-balance in distributed stream processing.
Our model integrated the distributed control message mechanism and a self-adapted
scheduler operator to detect hotspot issues quickly and perform state-migration
between operators efficiently. 
Experimental result shows that our model achieved better load balance on various
data-skewness workloads, and the hotspot operator is diminished according to its
latency and throughput. Our model also got better results under spikingly changing
situation due to the elastic time-window based algorithm.

In future work, we seek to adapt our AutoFlow model
to more scenarios like dynamic rescaling and automatic parallelization of dataflows.

\section{Acknowledgments}
  This work is supported by the National Key  R\&D Program of China under Grant No.2016YFB0200803; the National Natural Science Foundation of China under Grant No.61432018 and No.61602443; the Science Foundation of Beijing (L182053); Guangdong Province Key Lab of Popular High Performance Computers (2017B030314073) and the CAS Interdisciplinary Innovation Team of Efficient Space Weather Forecast Models.

\bibliographystyle{plain}
\bibliography{paper}

\vspace{12pt}

\end{document}